\documentstyle[a4wide,11pt]{article}
\title{Spacetime Structures and Physical Theories}
\author{Vu B Ho\\Department of Physics\\Monash University\\
Clayton Victoria 3168\\Austraila}
\date{}
\begin{document}
\maketitle
\begin{abstract}
General relativity is applied to the strong interaction; the nexus between the
two being arrived at by constructing a line element having the Yukawa form,
which is used to describe geometrically the classical dynamics of a particle
moving under the influence of the short-range strong interaction. 
It is shown that, with reasonable assumptions, the theory of general
relativity can be made compatible with quantum mechanics by using the
general relativistic field equations to construct a Robertson-Walker metric
for a quantum particle. The resulting line element of the particle can be
transformed entirely to that of a Minkowski spacetime, and the spacetime
dynamics of the particle described by a Minkowski observer takes the form
of quantum mechanics. It is also discussed the physical
aspects of the affine connection in general relativity and its relationship
with the field strength of the electromagnetic field and strong interaction. 
A heuristic geometric formulation of the electromagnetic field as an 
independent spacetime structure is presented. 
\end{abstract}

\newpage

\section{Introduction}

The geometrical character of physical reality has been discussed since the
time of the Greek philosophers. However, a serious
geometrial formulation of physical theories
only started with the advent of the general theory of relativity,
formulated by Einstein earlier this century \cite{Eins}.
Despite the fact that it is a relativistic generalisation of Newton's
theory of gravitation, general relativity is a theory about
curved spacetime structures in which gravity is described geometrically as
a manifestation of the curvature of a Riemannian spacetime manifold. The
dynamical description of physics then becomes force-free in the sense that
spacetime is also considered as a dynamical entity, whose metrical structure
is determined by matter fields through a system of nonlinear field
equations. The concept of geometrisation of physics has led to successful
theories in fundamental interactions in contemparary physics, especially
within the framework of gauge theories of elementary particles.
In this work we first consider the possibility of formulating the strong
interaction in terms of general relativity, focussing attention on the case
where the strong force between nucleons is described by the Yukawa
potential. Although general relativity was developed to describe the
gravitational field, its general formulation allows the formalism
to be applied to other physical fields, as long as the requirement of
mathematical consistency is satisfied. We apply the
formalism of general relativity to the `strong' interaction whose force
carriers are the Yukawa virtual pions. In this situation
the strong interaction is attractive and charge-independent,
although the matter that produces the attractive nucleonic force may not
have the same characteristics as the matter that produces the
gravitational field. It should be emphasised that gravitating matter
cannot be specified self-consistently (i.e., geometrically) within general
relativity, but can only be incorporated into the theory through the
energy-momentum tensor. General relativity may thus be viewed as
having the status of Newton's second law in classical mechanics, which
is regarded as a definition of force in terms of mass and acceleration. This
conceptual framework then allows for the expression of physical laws, such as
Newton's law of gravitation and the Lorentz force law of electromagnetism,
to be formulated \cite{Fey,Sym}. It is noted that the energy-momentum
tensor used to define the field equations of general relativity contains
only the general concept of force, and so it does not specify $\it
a~priori$ any particular physical law in nature. We verify the
applicability of the formalism of general relativity to the strong
interaction by showing that the field equations of general relativity
admit a line element that contains the Yukawa potential of the strong
force as an exact solution. This result reveals a possible relationship
between the strong interaction described in terms the Yukawa potential and
the mathematical formalism of general relativity.

The most interesting situation arises when we apply the general
relativistic formulation to quantum particles. When quantum
particles are described as spacetime structures, it is possible to make
general relativity compatible with quantum mechanics by constructing a
spacetime structure for a quantum particle, so that the dynamics of the
spacetime structure when viewed by a Minkowski observer will produce
quantum mechanics. In this case, space and time of the observer are
directly connected to, and determined by, the structure of space and time
of the quantum particle. The spacetime structure is determined by the
particle's own characteristics, such as energy density, and it is not
possible for an observer to specify in a deterministic way the physical
observables of the particle in the Minkowski spacetime. Furthermore, we
show that the geometrical and topological structures of a quantum particle
may be entirely different from that normally considered within
the present framework of quantum physics. For example, quantum particles
may exist as timeless four-dimensional objects. Such speculations on
quantum particles as hyperspheres have been discussed previously from a
philosophical viewpoint (see, e.g., \cite{Reich,Rucker}).

\section{General relativity and the strong interaction}
\subsection{Introductory remarks}

In this section we discuss a relationship between general relativity
and the strong interaction\footnote{By the strong interaction we are
referring to the nucleon-nucleon interaction mediating by Yukawa bosons
(i.e., pions).} which is assumed to be described by the Yukawa potential.
We consider this simple model because the form of the Yukawa potential can
be incorporated into a line element that satisfies the field equations of
general relativity. Since the strong interaction has been investigated
almost entirely within the context of a quantum theory, e.g , quantum
chromodynamics \cite{Ryde}, problems associated with the classical
dynamics of a particle under the influence of the strong force and the
concomitant classical laws of the strong force have not received much
attention. Obviously, a fundamental question is whether the strong
interaction also has a classical analogue? In classical physics, however,
since a particle is characterised only by its mass and charge, the dynamics
of a particle is assumed to follow either the laws of classical
electrodynamics or the laws of general relativity. Therefore, if the strong
force does not have any kind of relationship with the charge of a particle,
then it is conjectured that the dynamics of a particle under
the influence of the strong force would also follow the laws of general
relativity. This suggestion is based on the
fact that the strong interaction is attractive and charge-independent;
most importantly, the formalism of general relativity does not specify
{\it a~priori} the nature of matter that produces the effect of attraction.
Furthermore, the role of physical principles, such as the principle of
equivalence and the principle of covariance, incorporated into the general
relativistic formulation of the gravitational field can be argued to be
inessential (see, e.g., \cite{Syn}). Although the covariance principle
does make assertions about the mathematical formulation of the theory, it
does not have a direct physical significance in the sense that it is
possible to formulate covariantly any physical theory \cite{Paul}. The
status of the equivalence principle has also been subject to controvercy.
The problem here is related to whether it is physically significant to
specify a coordinate transformation to a local reference frame so that the
gravitational field is be eliminated \cite{Fock}. In the following, we
assume the applicability of the general relativistic formulation to the
strong interaction, and investigate the consequences that emerge from the
resulting formulation. Similar to the Newtonian gravitational field, that
forms a line element satisfying the field equations of general relativity,
we can construct a line element that contains an appropriate potential
characterising the strong force, and which satisfies the general
relativistic field equations. The relevant potential that we consider is
the Yukawa potential. This simple assumption and concomitant formulation,
therefore, does not take into account the quark structure of the nucleons
as in quantum chromodynamics (see, e.g., \cite{Aitc}). However, this
approach reveals a possible way to make general relativity compatible with
quantum mechanics, within the formulation of general relativity itself.

\subsection{A line element of the Yukawa potential}

In order to see whether the strong force can be describes by the laws of
general relativity, it is first necessary to consider whether the Yukawa
potential can form a line element that satisfies the field equations of
general relativity, $R_{\mu\nu}-\frac{1}{2}g_{\mu\nu}R+\Lambda g_{\mu\nu}
=\kappa T_{\mu\nu}$. Assuming a centrally symmetric field, the spacetime
metric can be written as \cite{Land}
\begin{equation}
ds^2=e^\mu c^2dt^2-e^\nu dr^2-r^2(d\theta^2+sin^2\theta d\phi^2).
\end{equation}
With this line element, the vacuum solutions satisfy the system
of equations
\begin{eqnarray}
\frac{\partial\mu}{\partial r}+\frac{1}{r}-\frac{e^\nu}{r}&=&0,\\
\frac{\partial\nu}{\partial r}-\frac{1}{r}+\frac{e^\nu}{r}&=&0,\\
\frac{\partial\nu}{\partial t}&=&0,
\end{eqnarray}
\begin{equation}
2\frac{\partial^2\mu}{\partial r^2}+\left(\frac{\partial\mu}{\partial r}
\right)^2+\frac{2}{r}\left(\frac{\partial\mu}{\partial r}-
\frac{\partial\nu}{\partial r}\right)-\frac{\partial\mu}{\partial r}
\frac{\partial\nu}{\partial r} -e^{\nu-\mu}\left[2\frac{\partial^2\nu}
{\partial t^2}+\left(\frac{\partial\nu}{\partial t}\right)^2 -
\frac{\partial\nu}{\partial t}\frac{\partial\mu}{\partial t}\right]=0.
\end{equation}
These equations are not independent, since it can be verified that the last
equation follows from the first three equations. Furthermore, the first two
equations give $\partial\nu/\partial r+\partial\mu/\partial r=0$ which
leads to $\nu+\mu=0$, due to the possibility of an arbitrary transformation
of the time coordinate.

If we assume a line element in the form of Yukawa potential, then it can be
written in the following simple form (see, e.g., \cite{Burc})
\begin{equation}
e^{-\nu}=1-\alpha\frac{e^{-\beta r}}{r},
\end{equation}
where the quantity $\alpha$ plays the role of the charge of a particle in
electromagnetism, and the quantity $\beta=1/R$, with $R=\hbar/mc$, specifies
the range of the strong force (which is assumed to be desc ibed by the
Yukawa potential). The quantity $m$ is the rest mass of Yukawa quanta,
(i.e., the virtual pions), whose continuous transfer between two nucleons
is assumed to give rise to the strong interaction. Since the quantity $\nu$
is now time-independent, by differentiating Eq.(6), it is found that
\begin{equation}
\frac{d\nu}{dr}=-\alpha\frac{e^{-\beta r}}{r}\frac{1+\beta r}{r -
\alpha e^{-\beta r}}.
\end{equation}
On the other hand, the quantity $d\nu/dr$ that follows from equations
(3) and (6) implies that
\begin{equation}
\frac{d\nu}{dr}=-\alpha\frac{e^{-\beta r}}{r}\frac{1}{r-\alpha e^{-\beta r}}.
\end{equation}
Equation (7) reduces to equation (8) if the condition $\beta r\ll 1$,
or $r\ll R$, is satisfied. In general, the quantity $R$ specifies a range,
so that the line element of the Yukawa form (6) can be approximated as
a solution to the
field equations of general relativity for the region $r\ll R$. It is seen
that for the maximal possible range, where $R\rightarrow\infty$, the metric
of the Yukawa form reduces to the familiar Schwarzschild metric, which is
used to describe a spherically symmetric gravitational field. In the case of
short range nuclear forces, the quantity $R$ can be assigned a value in
terms of the fundamental constants $\hbar$ and $c$, and the rest mass of the
Yukawa quanta. Consequently,  for the short range of the strong force,
the field equations of general relativity admit a line element that takes
the form of a Yukawa potential. This leads to the conclusion that by
specifying an appropriate matter source that characterises the strong
interaction, it is possible to consider the strong interaction as a
manifestation of general relativity at short range.

The strong interaction, however, has massive exchange quanta, unlike the
assumed massless force carriers of the gravitational field. The
masslessness of the latter quanta is consistent with the vacuum solution to
the field equations of general relativity. Therefore, any
solutions to the field equations of general relativity that are used to
describe the strong field should be non-vacuum solutions. In order to find
an appropriate solution to describe the strong force, an energy-momemtum
tensor must be specified. Obviously, at present nuclear physics does not
allow us to specify a presice form for the strong energy-momentum tensor. In
this situation it is appropriate to construct an energy-momentum tensor for
the strong force, so that it not only gives rise to the desired metric of
the Yukawa form as an exact solution to the field equations of general
relativity, but also satisfies the conservation law $T^\nu_{\mu;\nu}=0$.
In what follows we discuss a particular form of the strong energy momentum
tensor, which admits a line element of the Yukawa potential as an exact
solution to the field equations of general relativity. We consider a strong
energy-momentum tensor of the form
\begin{equation}
T_\mu^\nu = \left(\begin{array}{cccc}
-\frac{\alpha\beta}{\kappa}\frac{e^{-\beta r}}{r^2} & 0 & 0 & 0\\
0 & -\frac{\alpha\beta}{\kappa}\frac{e^{-\beta r}}{r^2} & 0 & 0\\
0 & 0 & \frac{\alpha\beta^2}{2\kappa}\frac{e^{-\beta r}}{r} & 0\\
0 & 0 & 0 & \frac{\alpha\beta^2}{2\kappa}\frac{e^{-\beta r}}{r}
\end{array}\right)
\end{equation}
With this energy momentum tensor, the field equations of general relativity
reduce to the system of equations
\begin{eqnarray}
e^{-\nu}\left(\frac{\partial\nu}{\partial r}-\frac{1}{r}\right)+\frac{1}{r}
&=&-\alpha\beta\frac{e^{-\beta r}}{r},\\
-e^{-\nu}\left(\frac{\partial\mu}{\partial r}+\frac{1}{r}\right)+\frac{1}{r}
&=&-\alpha\beta\frac{e^{-\beta r}}{r},\\
\frac{\partial\nu}{\partial t}&=&0,\\
-e^{-\nu}\left[2\frac{\partial^2\mu}{\partial r^2}+\left(\frac{\partial\mu}
{\partial r}\right)^2+\frac{2}{r}\left(\frac{\partial\mu}{\partial r}-
\frac{\partial\nu}{\partial r}\right)-\frac{\partial\mu}{\partial r}
\frac{\partial\nu}{\partial r}\right]&+&e^{-\mu}\left[2\frac{\partial^2\nu}
{\partial t^2}+
\left(\frac{\partial\nu}{\partial t}\right)^2-\frac{\partial\nu}{\partial t}
\frac{\partial\mu}{\partial t}\right]\nonumber\\
&=&\alpha\beta^2\frac{e^{-\beta r}}{r}.
\end{eqnarray}
As in the case of vacuum solutions, it can be verified that the first two
equations give $\partial\nu/\partial r+\partial\mu/\partial r=0$, which
leads to $\nu+\mu=0$; while equation (13) follows from the first three
equations. However, this system of equations when integrated gives a metric
of the following form \cite{Berg}
\begin{equation}
e^{-\nu}=1-\alpha\frac{e^{-\beta r}}{r}+\frac{Q}{r},
\end{equation}
where $Q$ is a constant of integration. The term $Q/r$ can be interpreted
as a Coulomb repulsive force which arises for charged particles,
such as two protons. This term, however, may be set to zero for strong
interactions that involve neutral particles, such as two neutrons.

The form of the energy-momentum tensor (9) is
mathematically valid since it can be shown to satisfy the conservation law
\begin{equation}
\nabla_\nu T^\nu_\mu=\frac{1}{\sqrt{-g}}\frac{\partial T^\nu_\mu\sqrt{-g}}
{\partial x^\nu} - \frac{1}{2}\frac{g_{\lambda\sigma}}{\partial x^\mu}
T^{\lambda\sigma}=0.
\end{equation}
An important feature emerges from the above model that relates to the nature
of the quantity $\alpha$ in the line element and the energy momentum
tensor. That is, since the quantities $\kappa$ and $\beta$ are positive,
the energy component $T_0^0$ and the quantity $\alpha$ always have opposite
signs. Therefore, since $g^{00}$ is positive,  if the energy component
$T_{00}$ is considered to be positive, then the energy component
$T^0_0=g^{00}T_{00}$ must also be positive, and in this case, the quantity
$\alpha$ must be negative. Since the quantity $\alpha$ in the line
element should be defined in terms of a matter source, the matter source
that produces the strong interaction is negative if the quantity $\alpha$
is negative. This property of matter will be discussed further later.
Furthermore, it is also noted from the metric of the Yukawa
potential that if the quantity $\alpha$ is negative, then there would be no
Schwarzschild-like singularity when the constant of integration $Q$ is set
to zero.

\subsection{Classical dynamics in the Yukawa strong force field}

If we assume that the motion of a particle in the Yukawa strong force field
is also governed by the geodesic equation
\begin{equation}
\frac{d^2x^\mu}{ds^2}+\Gamma^\mu_{\nu\sigma}\frac{dx^\nu}{ds}
\frac{dx^\sigma}{ds}=0,
\end{equation}
then using the metric with the Yukawa potential, i.e.,
\begin{equation}
ds^2=\left(1-\alpha\frac{e^{-\beta r}}{r}\right)c^2dt^2 - \left(1-
\alpha\frac{e^{-\beta r}}{r}\right)^{-1}dr^2 -
r^2\left(d\theta^2+sin^2\theta d\phi^2\right),
\end{equation}
the equations for the geodesics can be written explicitly as
\cite{Berg,Lawd}
\begin{equation}
\left(1-\alpha\frac{e^{-\beta r}}{r}\right)^{-1}\left(\frac{dr}{d\tau}
\right)^2 + r^2\left(\frac{d\theta}{d\tau}\right)^2 +
r^2sin^2\theta\left(\frac{d\phi}{d\tau}\right)^2 - c^2\left(1-\frac{\alpha
e^{-\beta r}}{r}\right)\left(\frac{dt}{d\tau}\right)^2=-c^2,
\end{equation}
\begin{eqnarray}
\frac{d}{d\tau}\left(r^2\frac{d\theta}{d\tau}\right) - r^2\sin\theta
\cos\theta\left(\frac{d\phi}{d\tau}\right)^2&=&0,\\
\frac{d}{d\tau}\left(r^2\sin^2\theta\frac{d\phi}{d\tau}\right)&=&0,\\
\frac{d}{d\tau}\left[\left(1-\alpha\frac{e^{-\beta r}}{r}\right)
\frac{dt}{d\tau}\right]&=&0.
\end{eqnarray}
By choosing spherical polar coordinates and considering motion in the
plane $\theta=\pi/2$, the equations (20) and (21) reduce to
\begin{equation}
\frac{d\phi}{d\tau}=\frac{l}{r^2}, \ \ \ \ \ \ \
\frac{dt}{d\tau}=\frac{kr}{r-\alpha e^{-\beta r}},
\end{equation}
where $l$ and $k$ are constants of integration. With these relations, the
equation for the orbit can be obtained from the equation (18) as
\begin{equation}
\left(\frac{l}{r^2}\frac{dr}{d\phi}\right)^2 + \frac{l^2}{r^2} =
c^2(k^2-1) + \frac{\alpha c^2}{r}e^{-\beta r} + \frac{\alpha l^2}{r^3}e^{-
\beta r}.
\end{equation}
In the case, when the condition $\beta r\ll 1$ is satisfied, and using the
approximation $e^{-\beta r}=1-\beta r$, the equation for the orbit becomes
\begin{equation}
\left(\frac{l}{r^2}\frac{dr}{d\phi}\right)^2 +
\frac{l^2(1+\alpha\beta)}{r^2} = c^2(k^2-(1+\alpha\beta)) + \frac{\alpha
c^2}{r} + \frac{\alpha l^2}{r^3}.
\end{equation}
By letting $u=1/r$ and differentiating the resulting equation with respect
to the variable $\phi$, it is found that
\begin{equation}
\frac{d^2u}{d\phi^2}+(1+\alpha\beta)u = \frac{\alpha c^2}{2l^2} +
\frac{3\alpha l^2}{2}u^2.
\end{equation}
Hence, the classical dynamics of a particle under the influence of strong
force of Yukawa potential is similar to that of a particle in the
Schwarzschild gravitational field. In the next section we discuss the quantum
dynamics of a particle in terms of general relativity, by showing that the
mathematical formulation of general relativity may contain within it the
quantum theory!

\section{General relativity and quantum mechanics}

It is often stated that general relativity may not be compatible with
quantum theory, because the former is formulated in terms of curved
spacetimes while the latter is formulated from the viewpoint of an observer
in flat Minkowski spacetime; the quantum dynamics of a particle is then
described in terms of a Hilbert space of physical states (see, e.g.,
\cite{Bohm}). Nevertheless, various attempts have been made to unify the
theory of general relativity with quantum theory by employing different
quantisation procedures \cite{Witt}. For example, the
path integral approach to quantum gravity has emerged as a promising
quantisation procedure \cite{Hawk}. However, by approaching the problem
in a more intuitive manner, we conjecture that general relativity and
quantum mechanics may be reconciled if the curved spacetime of a quantum
particle is constructed in such manner that it can be transformed to
Minkowski spacetime. As we discussed previously
it is possible to apply directly the general relativistic
formulation to the strong interaction so that a strong force exhibit a
geometrical character.

In the present section we are concerned with the possibility of deducing
the quantum dynamics of a particle from the mathematical formulation of
general relativity. First we note that despite the energy-momentum tensor
(9) gives an exact solution to the field equations of general
relativity, which, although mathematically acceptable, may not be
physically realistic. Nonetheless, the form of the energy-momentum tensor
(9) reveals the possibility that at the quantum level the energy density
may vary only as the inverse square of the distance, and that the pressure
may be ignored compared
to the energy density. With these assumptions, and for the purpose of
describing the spacetime dynamics of a single quantum particle, let us
consider a general relativistic spacetime model for a quantum particle
based on the Robertson-Walker metric (see, e.g., \cite{Narl,Peeb})
\begin{equation}
ds^2=c^2dt^2-S^2(t)\left(\frac{dr^2}{1-kr^2}+r^2(d\theta^2 + \sin^2\theta
d\phi^2)\right)
\end{equation}
and the energy-momentum tensor $T_{\mu\nu}$ of the form
\begin{equation}
T_\mu^\nu= \frac{A}{S^2}\left(\begin{array}{cccc}
1 & 0 & 0 & 0\\0 & 0 & 0 & 0\\0 & 0 & 0 & 0\\0 & 0 & 0 & 0
\end{array}\right).
\end{equation}
where $A$ is a constant. The quantity $S(t)$ is considered as the radius of
curvature and in general is a function of time which will be determined by
the field equations of general relativity. The parameter $k=-1,0,1$. The
sign of $k$ will be discussed shortly. 
Using the expression for the Ricci tensor in the form \cite{Narl}
\begin{equation}
R_{\mu\nu}=\frac{\partial^2\ln\sqrt{-g}}{\partial x^\mu\partial x^\nu} -
\frac{\partial \Gamma_{\mu\nu}^\sigma}{\partial x^\sigma} +
\Gamma_{\mu\sigma}^\lambda\Gamma_{\nu\lambda}^\sigma -
\Gamma_{\mu\nu}^\sigma\frac{\partial\ln\sqrt{-g}}{\partial x^\sigma},
\end{equation}
the field equations of general relativity then reduce to
\begin{eqnarray}
\frac{\dot{S}^2}{S^2}+\frac{kc^2}{S^2}-\frac{\Lambda c^2}{3}&=&\frac{\kappa
c^2}{3}\frac{A}{S^2}\\
2\frac{\ddot{S}}{S}+\frac{\dot{S}^2}{S^2}+\frac{kc^2}{S^2}-\Lambda c^2&=&0.
\end{eqnarray}
This system of equations has a static solution
\begin{equation}
S_0^2=\frac{kc^4}{4\pi G\epsilon}
\end{equation}
which is similar to the Einstein static model with an energy density
$\epsilon$ which may be very large \cite{Eins}. However, the
possibility of negative energy density should not be ruled out because at
the quantum level a particle may have a curved spacetime with negative
curvature as will be discussed presently. Since we are discussing curved
spacetimes at the quantum level, the quantity $\Lambda = k/S_0^2$ will
change drastically for a small variation in $S_0$. Hence, the quantity
$\Lambda$ may also be considered as an inverse square function of $S$,
i.e., $\Lambda = B/S^2$, where $B$ is constant\footnote{Actually the
constant $B$ could be set to zero in the discussion that follows.}.
The system of field equations (29) and (30) is then modified to the
following equations
\begin{eqnarray}
\frac{\dot{S}^2}{S^2}+\frac{kc^2}{S^2}-\frac{c^2}{3}\frac{B}{S^2}
&=&\frac{\kappa c^2}{3}\frac{A}{S^2}\\
2\frac{\ddot{S}}{S}+\frac{\dot{S}^2}{S^2}+\frac{kc^2}{S^2}-c^2\frac{B}{S^2}
&=&0.
\end{eqnarray}
This system of equations has a solution of the form
\begin{equation}
S=act, \ \ \ \ \ \ \mbox{where} \ \ \ \ \ a=\sqrt{\frac{\kappa A}{2}-k}
\end{equation}

Let us first consider the case $k=1$. It is seen that in this case a real
solution requires the spacetime of a quantum particle to have a very large
positive energy density in its own reference frame. The curved spacetime of
the particle in this case cannot be transformed to the Minkowski spacetime
of a quantum observer. However, results from nucleon scattering
experiments (see, e.g., \cite{Perk}) have shown that such a large energy
density is not appropriate for the surrounding spacetime of quantum
particles like protons and neutrons. Therefore, if we assume a reasonable
value for the energy density so that $\kappa A/2\ll 1$, then we are forced
to {\it quantize} the spacetime structure of the particle by introducing the
imaginary number $i$, hence $a\approx i$ or $S\approx ict$. The reason for
the introduction of this quantisation is that the Robertson-Walker metric
can be transformed to a manifestly Minkowski metric when $S=ict$.
Actually, this kind of quantisation turns the
pseudo-Riemannian
curved spacetime of the particle into a Riemannian spacetime. This means
that we assume the particle to be able to measure its temporal distance in
exactly the same way as its spatial distances. In other words, the particle
is viewed as an object existing as a timeless four-dimensional Riemannian
space. Speculations on elementary particles being tiny hyperspheres are
normally regarded as rather philosophical in nature. However, it is
interesting to quote \cite{Reich}, {\it `... Another case would arise
if space were four-(or more) dimensional in its smallest elements, but
three-dimensional as a whole. This situation would correspond to the case
of a thin layer of grains of sand which, although each is three-dimensional
if taken individually, taken as a whole forms essentially a two-dimensional
space. Similarly, atoms which individually are higher-dimensional could
cluster into three-dimensional structures. In such a world, a macroscopic
structure would have only the degrees of freedom of the three dimensions of
space, while an atom would have many more degrees of freedom. Sense
perceptions in such a world would not be noticeably different from those of
our ordinary world; and conversely, it is in principle possible to infer
from our ordinary experiences the higher-dimensional character of the
microscopic world. Incidentally, it is not impossible that quantum
mechanics will lead to such results.'}

Quantisation is realisable only when the Robertson-Walker metric (with
$S=ict$) of the particle can be transformed to a Minkowski metric. This
means that the curved spacetime structure of the particle can be viewed in
Minkowski spacetime.  This is in fact the case, for if we apply the
coordinate transformations \cite{Narl}
\begin{equation}
iR=ctr, \ \ \ \ \ cT=ct\sqrt{1-r^2}
\end{equation}
or
\begin{equation}
r=\left(1-\frac{c^2T^2}{R^2}\right)^{-1/2}, \ \ \ \ \ ct=iR\left(1-
\frac{c^2T^2}{R^2}\right)^{1/2}
\end{equation}
then, as can be readily verified, using the formula for the transformation
of the metric tensor
\begin{equation}
g'_{\mu\nu}=\frac{\partial x^\alpha}{\partial x^{'\mu}} \frac{\partial
x^\beta}{\partial x^{'\nu}}g_{\alpha\beta},
\end{equation}
the coordinate transformations (35) reduce the Robertson-Walker metric
(26) of the quantum particle (with $S=ict$) to a manifestly Minkowski
metric of the form
\begin{equation}
ds^2=c^2dT^2-dR^2-R^2(d\theta^2+\sin^2\theta d\phi^2).
\end{equation}
It is seen that the spacetime dynamics of the quantum particle
can now be investigated by an observer using a Minkowski metric. The
description can be achieved by writing the quantity $S$ in terms of
the coordinates $(R,cT)$ in the form of an action integral
\begin{equation}
S=-i\sqrt{c^2T^2-R^2}=-i\int ds=-ic\int\sqrt{1-\frac{v^2}{c^2}}dT,
\end{equation}
where $ds$ is the usual Minkowski spacetime interval and $v=R/T$. It should
be emphasised that because the above coordinate transformations involve
the imaginary number $i$, in order to get real observables of the quantum
particle, the velocity should be defined as $v=iR/T$. This definition is
consistent with the definition of the canonical momentum operator,
${\bf p}=-i\hbar\nabla$, in quantum mechanics. In order to relate this
result with the path integral formulation of quantum mechanics we introduce
a new quantity $\Psi$, defined by the relation $S=K\ln\Psi$, where
$K$ is a dimensional constant, which will be identified shortly. We then
obtain
\begin{equation}
\Psi= e^{\frac{i}{K}\int ds}.
\end{equation}
With this form, we can recover standard quantum mechanics, at least for
free particles in the nonrelativistic limit, by applying the Feynman path
integral method \cite{Feyn}.
Perhaps, the most important point that should be emphasised is that the
Minkowski coordinates in this case depend entirely on the metric structure
of the quantum particle. So observers in Minkowski spacetime cannot perform
measurements of physical observables of the particle using their own choice
of standard spatial and temporal gauges. This may be a reason for
the unpredictable behaviour of quantum particles formulated in Minkowski
spacetime. This leads to the conclusion that the action integral (39)
does not have a deterministic character, which is implied in its formulation
in classical physics. The intrinsic relationship between the spacetime of a
quantum particle and that of a Minkowski observer, as specified by the
transformations (26), also justifies the mathematical formulation of the
Feynman random path integral formulation of quantum mechanics.

In Minkowski spacetime the quantity $S$ has an action integral
form; in this case a quantum description can be constructed by following
Schr\"{o}dinger's original method \cite{Lud,Your}. We begin by noting
that the quantity $S$ satisfies the relation
\begin{equation}
-\frac{1}{c^2}\left(\frac{\partial S}{\partial T}\right)^2 + \left(
\frac{\partial S}{\partial R}\right)^2-1=0.
\end{equation}
Using the relation $S=K\ln\Psi$, the quantity $\Psi$ then satisfies the
following equation
\begin{equation}
-\frac{1}{c^2}\left(\frac{\partial\Psi}{\partial T}\right)^2 + \left(
\frac{\partial \Psi}{\partial R}\right)^2 - \frac{1}{K^2}\Psi^2=0.
\end{equation}

Since $\partial\Psi/\partial R=\nabla\Psi .(\partial{\bf R}/\partial R)
=|\nabla\Psi|\cos\alpha$, using the variational principle, after averaging
the above equation with $\left<\cos^2\alpha\right>=1/2$, we obtain a
Klein-Gordon-like wave equation
\begin{equation}
-\frac{1}{c_a^2}\frac{\partial^2 \Psi}{\partial T^2}+\nabla^2\Psi -
\frac{1}{K_a^2}\Psi=0,
\end{equation}
where $c_a=c/\sqrt{2}$ and $K_a=K/\sqrt{2}$. If this equation is compared
with the Klein-Gordon equation in relativistic quantum mechanics
(\cite{Mess}), then it is seen that equation (43) describes the quantum
dynamics of a particle with an average velocity $c_a$ which is less than
the velocity of light $c$. This may give a reason why the force carriers in
the strong and weak interactions are massive. The comparison between
Eq.(43) and the Klein-Gordon equation also gives $K_a=\hbar/mc_a$.
We remark that the treatment of a quantum particle as a spacetime
manifold possessing a Robertson-Walker metric is consistent with our
previous static solution in which the Yukawa potential was used to
construct a line element for the strong interaction. This is because the
Yukawa potential is actually derived from the Klein-Gordon wave equation in
relativistic quantum mechanics.

Now let us consider the case $k=-1$. In this case we have $a\approx 1$
or $S\approx ct$, again assuming $\kappa A/2\ll 1$. This is in fact the well
known Milne model that arises from Milne's work on kinematic relativity
\cite{Milne}. The coordinate transformations of the form \cite{Narl}
\begin{equation}
R=ctr, \ \ \ \ \ cT=ct\sqrt{1+r^2}
\end{equation}
or
\begin{equation}
r=\left(\frac{c^2T^2}{R^2}-1\right)^{-1/2}, \ \ \ \
ct=R\left(\frac{c^2T^2}{R^2}-1\right)^{1/2}
\end{equation}
also reduce the Robertson-Walker line element (26) of a quantum particle,
with $S=ct$, to that of Minkowski spacetime. The quantity $S$ written in
terms of the coordinates $(R,cT)$ as a classical action integral now takes
the form
\begin{equation}
S=\sqrt{c^2T^2-R^2}=\int ds=c\int\sqrt{1-\frac{v^2}{c^2}}dT
\end{equation}
and satisfies the relation
\begin{equation}
-\frac{1}{c^2}\left(\frac{\partial S}{\partial T}\right)^2 + \left(
\frac{\partial S}{\partial R}\right)^2+1=0.
\end{equation}
The quantity $\Psi$, defined by the relation $S=K\ln\Psi$, now becomes
\begin{equation}
\Psi=e^{\frac{1}{K}\int ds}.
\end{equation}
Using the variational principle, it can be shown that $\Psi$ satisfies the
equation
\begin{equation}
-\frac{1}{c_a^2}\frac{\partial^2 \Psi}{\partial T^2}+\nabla^2\Psi +
\frac{1}{K_a^2}\Psi=0.
\end{equation}
Equation (49), however, differs from the Klein-Gordon equation by the
appearance of a plus sign in the last term. This results from the fact that
the quantum particle in this case has negative curvature. This kind of
particle structure is not usually found in quantum mechanics. However, it is
seen that if the energy density is negative, then our
formulation is real-valued and compatible with the usual relativistic
description with timelike intervals. If the spacetime structure of a
quantum particle in this case is also {\it quantised}, by turning equation
(49) into the Klein-Gordon equation, so that the `conventional'
quantum mechanics is used to describe the quantum dynamics of the particle,
then we replace $K\rightarrow iK$. This process of quantisation is
equivalent to turning the spacetime structure of a particle with negative
curvature into that with positive curvature, and specifying a positive
energy density for the particle in Minkowski spacetime.

Finally, in the case when $k=0$, a real solution is obtained for any
positive energy density, $S=act$. The spatial part of the Robertson-Walker
line element simply becomes the Euclidean metric scaled by the factor $S$.
If we apply the coordinate transformations
\begin{equation}
R=actr, \ \ \ \ \ \ cT=ct
\end{equation}
then $dR=actdr+acrdT$. It is seen that when the term $acrdT\ll 1$, the
spacetime structure of a particle can be reduced to that of the Minkowski
spacetime. For a particle with large energy density, i.e., $a$ large, its
curved spacetime metric can only be transformed to a Minkowski metric for
a short time $dT$.

We conclude this section with some remarks on the
energy-momentum conservation laws in general relativity. For the simple
models that have been discussed the solution $S=S_0$ satisfies strictly the
conservation laws required by general relativity. However, when the theory
is applied to quantum physics, the conservation laws should not be expected
to be satisfied strictly at the quantum level; the uncertainty
principle would allow violations. Furthermore, it seems that at the quantum
level constraints such as positivity of energy density also become relative
and coordinate-dependent, and this may affect the way in which a Minkowski
observer describes a physical process. These fundamental problems are as
yet unresolved and require further investigation.

\section{Further discussions and speculations}
\subsection{A heuristic geometric formulation of electromagnetism}

We first remark that all attempts at the unification of
the electromagnetic field with the gravitational field have focused on
incorporating electromagnetism into gravitation by modifying a Riemannian
metric structure; this metric structure essentially describes the
gravitational field. Since all such attempts have not been
successful, a natural question arises as to whether electromagnetism
should be regarded as more fundamental than gravitation, so that
electromagnetism should be described geometrically as an independent affine
structure of a spacetime manifold, without referring to the existence of
a Riemannian metric structure that represents the gravitational field. This
radical alternative approach to the geometrisation of physics may allow
gravity to arise as an additional structure to the spacetime
structure of electromagnetism. In this section we discuss
how a heuristic geometrical formulation of electromagnetism can be realised
in terms of non-Riemannian geometry. In this formulation, a simple
quantisation procedure can be introduced into an electromagnetic
structure of the spacetime manifold to make it compatible with the quantum
theory. This quantisation is carried out by considering complex-valued
changes to a vector under an infinitesimal parallel displacement. In this
case, there are similarities between the formulation of quantum mechanics
in an electromagnetic spacetime and the theory of gravitation using the
Riemannian spacetime. The free particle Schr\"{o}dinger
wave equation in a Euclidean space must be modified to the free particle
Schr\"{o}dinger wave equation in an electromagnetic spacetime. The latter
wave equation is identical to the wave equation of a charged particle moving
in an electromagnetic field in the background Euclidean space.

We adopt the view that electromgnetism should be described in its own
right, as an independent affine structure of a spacetime manifold without
reference to any other possible metric structures, such as the Riemannian
metric structure used to describe the gravitational field. This approach is
equivalent to postulating independent physical fields in field
theory where, with the same background Minkowski spacetime, a particular
field can be formulated by the introduction of a particular mathematical
structrue.  It is assumed that a spacetime manifold can be endowed with
a geometric structure from which a particular, affine or metric, structure
can be chosen to describe a physical field. Unless it is equipped with a
geometric structure, the background spacetime manifold cannot describe
physical laws.

In order to describe electrodynamics geometrically, an affine connection is
introduced into the differentiable manifold of spacetime. The introduction
of such a connection can be carried out by adopting a heuristic approach
modelled on parallel transport of a vector field and its covariant
derivative. Instead of the form given to the connection used to
formulate general relativity, the change $\delta V^\mu$ in the components
of a vector $V^\mu$ under an infinitesimal parallel displacement in the
present situation is assumed to be of the form
\begin{equation}
\delta V^\mu = -\beta \Lambda_\nu V^\mu dx^\nu,
\end{equation}
in which case the covariant derivatives are defined as
\begin{eqnarray}
\nabla_\nu V^\mu&=&\frac{\partial V^\mu}{\partial x^\nu} +
\beta \Lambda_\nu V^\mu,\\
\nabla_\nu V_\mu&=&\frac{\partial V_\mu}{\partial x^\nu} -
\beta \Lambda_\nu V_\mu.
\end{eqnarray}
These are required to transform like a tensor under general coordinate
transformations. Here the quantity $\Lambda_\nu$ is an affine connection
of the spacetime manifold, which will be identified with the electromagnetic
four-vector potential. The quantity $\beta$ is an arbitrary dimensional
constant. The transformation law for the affine connection $\Lambda_\mu$ can
be deduced from the transformation properties of the covariant derivative.
Under a general coordinate transformation $x'^\mu=x'_\mu(x^\nu)$, the
connection $\Lambda_\mu$ transforms as
\begin{equation}
\beta{\Lambda'}_\mu = \frac{\partial x^\nu}{\partial {x'}^\mu} (\beta
\Lambda_\nu) + \frac{\partial^2 x^\nu}{{\partial {x'}^\mu}{\partial
{x'}^\sigma}} \frac{\partial {x'}^\sigma}{\partial x^\nu}.
\end{equation}

The generalisation of Eqs.(52) and (53) can be obtained from the
definition of a covariant derivative, i.e.,
\begin{eqnarray}
\nabla_\sigma A^{\mu_1...\mu_m}&=& \frac{\partial A^{\mu_1...\mu_m}}
{\partial x^\sigma}+m\beta\Lambda_\sigma A^{\mu_1...\mu_m},\\
\nabla_\sigma A_{\nu_1...\nu_n}&=&\frac{\partial A_{\nu_1...\nu_n}}
{\partial x^\sigma}-n\beta\Lambda_\sigma A_{\nu_1...\nu_n},\\
\nabla_\sigma A^{\mu_1...\mu_m}_{\nu_1...\nu_n} &=& \frac{\partial
A^{\mu_1...\mu_m}_{\nu_1...\nu_n}}{\partial x^\sigma} + (m-n)\beta
\Lambda_\sigma A^{\mu_1...\mu_m}_{\nu_1...\nu_n}.
\end{eqnarray}
It is interesting to note that the covariant derivative of a mixed tensor
having equal number of superscripts and subscripts is identical to its
ordinary derivative.

It should be reiterated that the electromagnetic structure is assumed to
be an independent structure, which defines a curved spacetime.
This must not be considered as an additional structure arising from the
postulate of gauge invariance, as in Weyl's theory, which assumes a change
$\delta l=-l\phi_\mu dx^\mu$ of the length $l=g_{\mu\nu}\xi^\mu\xi^\nu$ of a
vector $\xi^\mu$ under parallel transport. In this latter case $g_{\mu\nu}$
represent a Riemannian spacetime structure of gravitation and $\phi_\mu$ is
a four-vector function which is identified with the
four-vector potential of an electromagnetic field. The electromagnetic
`spacetime' in our case is assumed to exist by itself, independent of any
other spacetime structures, such as the gravitational field. The purpose of
the introduction of the connection $\Lambda_\mu$ is to construct a
{\it non-Riemannian} spacetime manifold
which can be used to represent electromagnetism alone. In this way
that an appropriate topological structure of the manifold can be related
to the quantum dynamics of a particle in spacetime.

The quantities $\Lambda_\mu$ in general are arbitrary functions of the
coordinate variables and they do not form a tensor under general coordinate
transformations, but they do form a tensor under the group of linear
transformations. This result is compatible with the usual formulation of
electromagnetism as a physical field in a Minkowski spacetime, which is
invariant under Lorentz transformations. In a particular coordinate system,
if the connection is identified with the electromagnetic potentials, then,
since the four-potential should transform as a vector, the effect caused by
the extra term in the transformed connection (54) is not an electromagnetic
effect. It is related purely to coordinate transformations, i.e., it
represents an {\it inertial effect}. Therefore, if the potentials are
significant then our geometrical formulation of electromagnetism is
covariant only under the group of linear tranformations. This is reasonable
since an inertial effect caused by non-linear coordinate transformations
has not been related to any kind of eletromagnetic properties. The
formulation of a physical theory is normally required to be covariant only
under some particular group of transformations, except for the general
theory of relativity in which the formalism is based on the requirement of
general covariance \cite{Eins}.

However, in the present geometrical formulation of electromagnetism,
the geometrical object which plays the role of the Riemannian curvature
tensor is covariant under general coordinate transformations. This object
is expected to take the familiar form of the electromagnetic field tensor
$F_{\mu\nu} = \partial_\mu \Lambda_\nu -\partial_\nu \Lambda_\mu$.
This curvature can be derived by considering the change
$\Delta V_\mu=\oint\delta V_\mu$ of a vector $V_\mu$ parallel transported
around an infinitesimal closed path. To first order, an infinitesimal closed
path permits the components of the vector $V_\mu$ at points inside the path
to be uniquely determined by their values on the path. By Stokes theorem, it
is found that \cite{Land}
\begin{eqnarray}
\Delta V_\sigma &=& \oint_\Gamma \Lambda_\nu V_\sigma dx^\nu\nonumber\\
&=& \frac{1}{2}\left (\frac{\partial \Lambda_\mu}{\partial x^\nu}-
\frac{\partial \Lambda_\nu}{\partial x^\mu}\right) V_\sigma\Delta
f^{\mu\nu},
\end{eqnarray}
where $\Delta f^{\mu\nu}$ is the area enclosed by the closed path $\Gamma$.
Since $V_\mu$ is a vector and $\Delta f^{\mu\nu}$ is a tensor, and since
$\Delta V_\mu$ is
also a vector, because it is the difference between the values of vectors
at the same point after parallel displacement, the tensor character of
the curvature, defined by the relation $ F_{\mu\nu}=\partial_\mu\Lambda_\nu
-\partial_\nu\Lambda_\mu$ is determined from the quotient theorem in tensor
calculus. The quantities $F_{\mu\nu}$ therefore form a tensor under
general coordinate transformations. This result shows that if only the
field strength of the electromagnetic field is considered significant, then
the present geometrical formulation of electromagnetism, like the general
relativistic formalism of gravitation, is also covariant with respect to
the general group of transformations. However, as we shall discuss in detail
later, when the electromagnetic field strength is defined in terms of the
potentials, the existence of the electromagnetic field strength
requires restrictions on the analytic properties of the four-potential.
We shall also demonstrate that the present formulation of
electromagnetism may lead to the possibility of introducing an asymmetric
connection, in which the combined electromagnetic effects of two
electromagnetic fields on a charged particle provide a geometrical framework
for describing the dynamics of the particle. Remarkably, we can show that
when the combined effects on a charged particle are electromagnetically
neutral\footnote{Electromagnetic neutrality is defined within the context
of the Lorentz force law, where opposing electromagnetic fields lead to the
possibility of the net force on the particle being zero.},
they can be identified with gravity within a general relativistic framework.

The curvature $F_{\mu\nu}$ automatically satisfies the homogeneous
equations of classical electrodynamics, $\partial_\alpha F_{\mu\nu}+
\partial_\mu F_{\nu\alpha} + \partial_\nu F_{\alpha\mu}=0$.
The result shows that the homogeneous equations of electrodynamics are
geometrical rather than dynamical when the connection $\Lambda_\mu$ is
considered as being a purely geometrical object. As usual, to determine the
dynamics of the electromagnetic spacetime manifold, an action, that
may or may not relate geometrical properties of the manifold to matter or
charge, must be specified. If such an action is defined by the form $S =
-\int\left( F_{\mu\nu} F^{\mu\nu}-\kappa\Lambda_\mu j^\mu \right) dx^4$,
where $\kappa$ is an arbitrary dimensional constant, then the variation of
the action $S$ with respect to the connection $\Lambda_\mu$ leads to the
inhomogeneous equations of classical electrodynamics, i.e., $\partial_\mu
F^{\mu\nu}+\kappa j^\nu=0$. The external current density, $j^\mu$, whose
geometrical character is unknown, plays the role of the stress tensor in the
field equations of general relativity. In the case when there is no external
current, the field equations of electromagnetism $\partial_\mu F^{\mu\nu}=0$
describe a vacuum spacetime structure, similar to the source free
gravitational field equations, $R_{\mu\nu}=0$, in general relativity.

The spacetime structure of electromagnetism that has been described using
the connection $\Lambda_\mu$ and the curvature $F_{\mu\nu}$ is entirely
affine. An affine structure is not capable of providing a dynamical
description of the motion of a particle in the field. This is exactly the
case in classical electrodynamics where the Lorentz force must be added to
the Maxwell field equations for a dynamical description of a charged
particle. With a geometrical formulation of the physical field, the
dynamics can be provided by introducing a metric tensor $g_{\mu\nu}$
onto the spacetime manifold through the defining relation $ds^2 =
g_{\mu\nu}dx^\mu dx^\nu$. When the spacetime manifold is endowed with a
metric, a relationship between the metric and the connection can be obtained
by demanding that the metric be covariantly constant, in the sense that
the inner product of two vectors remains constant under parallel transport
along a curve. This requirement leads to the condition
\begin{equation}
\nabla_\sigma g_{\mu\nu}=\frac{\partial g_{\mu\nu}}{\partial x^\sigma}-
2\beta\Lambda_\sigma g_{\mu\nu}=0.
\end{equation}
Eq.(59) can be rewritten in the form
\begin{equation}
\Lambda_\mu = \frac{g^{\lambda\sigma}}{2\beta} \frac{\partial
g_{\lambda\sigma}}{\partial x^\mu} =
\frac{1}{2\beta g}\frac{\partial g}{\partial x^\mu} =
\frac{1}{\beta} \frac{\partial \ln\sqrt{-g}}{\partial x^\mu},
\end{equation}
where $g=\det(g_{\mu\nu})$. Under the usual gauge transformation
\begin{equation}
{\Lambda'}_\mu = \Lambda_\mu + \frac{\partial \chi}{\partial x^\mu},
\end{equation}
with $\chi=(\ln\sqrt{\sigma})/\beta$, we obtain
\begin{equation}
{\Lambda'}_\mu = \frac{1}{\beta} \frac{\partial
\ln\sqrt{-\sigma g}}{\partial x^\mu} =\frac{1}{2\beta} \frac{1}{(\sigma g)}
\frac{\partial {(\sigma g)}}{\partial x^\mu}.
\end{equation}
Here $\chi$, and hence $\sigma$, is an arbitrary function of the
coordinate variables.

If the potentials are considered to have a physical significance and
covariance of the theory is imposed, the transformation group is restricted
to the linear group. In this case it is not possible to introduce a
metric tensor that satisfies the requirement of being covariantly constant
$\nabla_\sigma g_{\mu\nu}=0$, since linear coordinate transformations
imply $g=\det(g_{\mu\nu})=constant$ and the affine connection $\Lambda_\mu$
defined in terms of the metric tensor in Eq.(60) vanishes. This
result leads to the conclusion that electromagnetism is a non-metrical
spacetime structure. Although the conclusion seems to contradict a
basic requirement of classical physics, namely, that the local validity of
physical laws requires a metric, the non-metric character of
electromagnetism seems to comply with the principles of quantum mechanics
whose non-deterministic formulation implies the impossibility of using a
metric for investigating of the dynamics of a physical system.
In other words, the process of physical measurement in a deterministic
manner is not possible for the electromagnetic field, which is covariant
only under the linear group of coordinate transformations.

On the other hand, if the field strength is considered to be the relevant
physically significant object, as in classical electrodynamics, and the
effects caused by non-linear
coordinate transformations are treated as inertial effects, then a metric
tensor can be introduced so that the electromagnetic potential can be
defined through it. However, the relationship between the connection, the
curvature and the metric tensor means that the electromagnetic field
strength can exist only at spacetime points where the condition of
integrability is not satisfied. In spacetime regions where the determinant
of the metric tensor is a smooth function, or at least twice
differentiable, there will only be a four-potential without a concomitant
electromagnetic field strength. It is the analytical behaviour of the
metric tensor that determines the existence of the electromagnetic field
strength. This result leads to the conclusion that for metric spacetime
structures describing electromagnetism, for which the electromagnetic
potentials are considered significant, it is not possible to measure the
electromagnetic field strength of the system smoothly (i.e., the field
strength is not differentiable). This clearly contradicts the usual
assumption that physical laws can be formulated locally in classical
physics. Whether this situation can be incorporated consistently
into the present formulation of quantum mechancis is a problem that
requires further investigation. For example, it is known that quantum
mechanics utilises potentials to describe the Aharonov-Bohm effect.

The equation of motion of a charged particle in an electromagnetic spacetime
manifold can also be obtained from the requirement that the path of a
particle is a geodesic, i.e.,
\begin{equation}
\frac{d^2 x^\mu}{ds^2} + \beta \Lambda_\nu \frac{dx^\nu}{ds}
\frac{dx^\mu}{ds} = 0.
\end{equation}
where the parameter $s$ is identified with the arc-length only when a
metric exists. Since the affine connection $\Lambda_\mu$ is entirely
geometrical, the equation of motion in this form does not have the dynamical
character required of a physical process, where some physical quantity is
required to characterise the physical state of a particle. Therefore, it is
necessary to introduce some kind of relationship between the geometrical
objects and the physical quantities in order to provide a possible
dynamical description of the system consisting of particle and field. For
example, if the following relationship is assumed
\begin{equation}
\beta \Lambda_\nu \frac{dx^\mu}{ds} = -\frac{q}{m} F^\mu_\nu,
\end{equation}
then the familiar form of Lorentz force law for the motion of a charged
particle in an electromagnetic field is regained, i.e.,
\begin{equation}
\frac{d^2x^\mu}{ds^2} = \frac{q}{m}F^\mu_\nu \frac{dx^\nu}{ds}.
\end{equation}
However, since there is no physical basis for its introduction, the
relation (64) should be considered as an intrinsic relationship between
the field and the experimentally defined physical quantities that
characterise the mass and the charge of a particle.

In the case when a metric is introduced onto the electromagnetic manifold,
the relation (60) between the metric tensor and the connection can be
rewritten in the form
\begin{equation}
g=g_0\exp\left(2\beta\int\Lambda_\mu dx^\mu\right).
\end{equation}
In the non-relativistic limit, the determinant $g$ can be reduced further
to the form
\begin{equation}
g=g_0\exp\left(2\alpha\frac{ct}{r}\right),
\end{equation}
where $g_0$ and $\alpha$ are constants. Utilising this form of the
determinant of the metric tensor, with $\Lambda_\mu=(\phi,-{\bf A})$, the
four-vector potential is obtained
\begin{equation}
\phi = \frac{\alpha}{\beta}\frac{1}{r}, \ \ \ \
{\bf A} = \frac{\alpha ct}{\beta} \frac{{\bf r}}{r^3}.
\end{equation}
With this form of the four-potential, it is noted that at finite time $t$,
except at the origin $r=0$, the electromagnetic field strength defined
in terms of the four-potential vanishes everywhere. Any possible influence
on a charged particle in this electromagnetic spacetime
can only be expressed in terms of the potentials themselves. This result is
similar to the situation in general relativity where the effect of a
gravitational field on a particle can only be expressed in terms of the
potential, which forms a line element for the field.

The non-relativistic equation of motion takes the form
\begin{equation}
\frac{d^2 {\bf r}}{dt^2} + \frac{c\alpha}{r}\frac{d{\bf r}}{dt} = 0,
\end{equation}
where the scalar potential $\phi=\alpha/\beta r$ has been used. It can be 
shown that a semi-classical relation can be derived
from differential geometry which relates the momentum and the de Broglie
wavelength of a particle, i.e.,
\begin{equation}
\frac{d{\bf r}}{dt}=\frac{\hbar}{m}\frac{{\bf r}}{r^2}.
\end{equation}
The equation of motion then takes the familiar form of Newton's equation of
motion for a charged test particle moving in a spherically symmetric
electrostatic field
\begin{equation}
m\frac{d^2 {\bf r}}{dt^2} = -e^2 \frac{{\bf r}}{r^3}.
\end{equation}
Here we have set $\alpha=e^2/\hbar c$, with $e$ denoting the fundamental
electronic charge. The constant $\alpha$ is the fine structure constant,
and the constant $\beta$ is equal to $1/\hbar c$. In this case the electric
field strength is determined experimentally, since classically the vector
potential $\bf A$ is considered merely as a mathematical convenience, with
no direct physical significance, and so would have no physical effect on
the motion of a charged particle (see, e.g., \cite{Jackson}).

Let us now consider a simple procedure for quantising the electromagnetic
spacetime structure. It is known that the canonical quantisation of a
classical field requires the introduction of the imaginary number $\it i$
into the mathematical structure used to formulate the theory in an
appropriate way. Specifically, quantisation is achieved by replacing the
classical Poisson bracket by the quantum Poisson bracket; namely,
$[q,p]\rightarrow i\hbar[q,p]$. Alternative quantisation procedures, such
as the path integral formulation of quantum mechanics, introduce a
complex-valued transition amplitude; however, this method still retains the
classical notion of an action integral. We discussed previously a possible
relationship between the theory of general relativity and quantum theory,
in which it is necessary to employ the quantity $\it i$ in order to
obtain a coordinate transformation from the spacetime structure of a quantum
particle to that of a Minkowski observer. In so doing the spacetime dynamics
of a quantum particle manifests itself in Minkowski spacetime as the
quantum dynamics described by a relativistic wave equation. We now examine
the quantisation procedure within the present geometric formulation of
the electromagnetic field. It has already been noted that the determinant
of a metric tensor on an electromagnetic spacetime manifold has the property
that it generates an electromagnetic field through its analytical behaviour.
This property is reflected in the gauge transformation (62). Each quantity
$g$ corresponds to an infinite number of possible spacetime metrics, with
the same electromagnetic field generated by all analytically equivalent
determinants of the form $\sigma g$. For example, there is an infinite
number of metric structures corresponding to the potentials given in
(68). Within the context of classical electrodynamics, the vector
potential ${\bf A}=(\alpha ct{\bf r})/(\beta r^3)$ in this case produces
a zero magnetic field, i.e., ${\bf B}=\nabla\times {\bf A}\equiv 0$. On the
other hand, the scalar potential $\alpha/(\beta r)$ can be identified with
the Coulomb potential used in the quantum dynamical description of a hydrogen
atom. Therefore, a dynamical equation, such as the Schr\"{o}dinger wave
equation, that contains only the potentials (68) cannot be used to
determine the metric structure of a physical system. This implies the
impossibility of determining the path of a particle. Let
us now consider a purely imaginary change of a vector under an infinitesimal
displacement
\begin{equation}
\delta A^\mu = -i\beta \Lambda_\nu A^\mu dx^\nu.
\end{equation}
The covariant derivative will take the form
\begin{equation}
\nabla_\nu A^\mu =\frac{\partial A^\mu}{\partial x^\nu} +
i\beta \Lambda_\nu A^\mu.
\end{equation}
It is known that the gauge invariant formulation of quantum
mechanics requires the free particle Schr\"{o}dinger wave equation
to be modified when analysing a charged particle moving in an
electromagnetic field. In both cases the free particle equation
and the modified equation take the Euclidean space as their background
space. We can demonstrate that this situation may be described
geometrically within the present formulation of electromagnetism. First we
recall that general relativity describes the gravitational field
in terms of geometry, without the concept of force; namely, the free
particle Newtonian equation, $\eta_{\mu\nu}(dx^\mu/ds)(dx^\nu/ds)=0$, in a
Euclidean space must be modified to the geodesic equation,
$d^2x^\mu/ds^2+\Gamma^\mu_{\sigma\lambda} (dx^\sigma/ds)(dx^\lambda/ds)=0$,
in a Riemannian manifold to describe the dynamics of a particle in a
gravitational field. In the present case, however, if we take the
Schr\"{o}dinger wave equation as the fundamental dynamical equation, then
the free particle Schr\"{o}dinger wave equation, $i\hbar\partial_t\psi +
(\hbar^2/2m)\nabla^2\psi=0$, in a Euclidean space must be replaced by the
wave equation, $i\hbar\nabla_0\psi + (\hbar^2/2m)\nabla_\mu
\nabla^\mu\psi=0$, in a curved electromagnetic manifold for the dynamical
description of a charged particle in an electromagnetic field. However, the
problem that remains here is whether it is possible to derive the
Schr\"{o}dinger wave equation from the field equations of the
electromagnetic spacetime manifold, as is the case in general relativity
where the geodesic equation for a particle in a gravitational field can be
derived from the Einstein field equations. However, as we already showed,
it is indeed possible to obtain the Schr\"{o}dinger wave equation from the
field equations of general relativity by a coordinate transformation.

\subsection{Physical aspects of the affine connection}

General relativity is a physical theory that describes particle dynamics
in terms of the geometry of a spacetime manifold. In this interpretation,
gravity is a manifestation of the curvature of a Riemannian metric spacetime
structure under the influence of matter. In Einstein's theory of
gravitation, the metric tensor plays the role of the gravitational potential,
while the metric connection, expressed in terms of the metric tensor, only
plays an intermediate role and has no direct physical interpretation.
However, the more fundamental object in the differential geometric
formulation of spacetime manifolds is the affine connection. The question
arises as to whether it is possible to assign a direct physical meaning to
the connection, in the sense
that the connection will play the fundamental role of a physical field
which actually determines the dynamics of a particle. This point of view can
be justified in part by noting that the geodesic equation of motion
is directly governed by the connection itself. Historically, in an attempt
to describe gravitation and electromagnetism within the context of
a single spacetime structure, Einstein tried to formulate a `unified' theory
using a nonsymmetric affine connection rather than the metric tensor
\cite{Eins}.

Unlike other equations in physics, which are usually deduced from
experiments, and formulated as relations between physical quantities,
the field equations of general relativity, $R_{\mu\nu}-\frac{1}{2}
g_{\mu\nu}R=\kappa T_{\mu\nu}$, were postulated via an examination of tensor
properties of the Einstein tensor, and the stress tensor. The former is a
purely geometrical object, while the latter is a physical quantity that has
no direct mathematical
interpretation as a geometrical object. The question has been raised as to
whether it is possible to construct a purely geometrical model for the
theory of general relativity by looking for a geometrical interpretation
of the stress tensor \cite{Gibb}. However, an alternative viewpoint can
be adopted in which we look instead for a physical construction of the
Einstein tensor in terms of known physical entities, so that the field
equations of general relativity adopt the status of physical equations,
describing a general relationship between purely physical quantities.
This alternative
viewpoint provides the basis for a physical interpretation of the affine
connection in the theory of general relativity. In what follow we discuss
the possibility of constructing the affine connection in terms of the
potentials and the field strengths of two coupled electromagnetic fields.

In classical physics, electromagnetism and gravitation are considered to be
two different physical structures that exist independently of each other in
the spacetime manifold. In such a situation a gravitational field is not
considered to have a direct relationship with the charge of a particle in
the sense that there is no interaction between them. On the other hand,
the mass of the particle is not related to the electromagnetic properties of
an electromagnetic field. Only when the dynamics of the particle is studied
using the laws of motion are these dynamical aspects connected. Consider a
charged particle in a region of spacetime in which there is an
electromagnetic field $\Phi_\nu=(\phi, -{\bf A})$, where $\phi$ and
${\bf A}$ are the scalar and vector potential, respectively. If we assume
that in this same region another electromagnetic field could be set up so
that at every point in the region the magnitudes of the two fields are
always equal but the two fields are in opposite directions, then, according
to the classical Lorentz force law, the charged particle would not be
affected electromagnetically by the fields because their effects on the
particle cancel. In this case the existence of the electromagnetic
fields can be ignored within the context of classical electrodynamics. In
other words, the spacetime surrounding region is `neutralised' by the
presence of the second field. The spacetime region surrounding an hydrogen
atom, or the spacetime region surrounding the earth, obviously satisfies
this requirement. For example, if a particle $A$ with charge $q$ and
a particle $B$ with charge $-q$ are located at the origin of a coordinate
system, then the field strengths and the potentials of the particles are
\begin{eqnarray}
{\bf E}_A &=&\frac{kq}{r^3}{\bf r}, \ \ \ \ \ \ \phi_A=-\frac{kq}{r}\\
{\bf E}_B &=&-\frac{kq}{r^3}{\bf r}, \ \ \ \ \ \phi_B=\frac{kq}{r},
\end{eqnarray}
and Newton's law of motion $m{\bf a}={\bf F}$ would result in a net zero
effect on a charged particle in the surrounding spacetime region, since
in classical physics, the potentials play no direct role in determining
the dynamics of a particle but are merely a mathematical convenience
\cite{Jackson}. 
However, this situation changes if quantum effects are taken into account,
since it is known that in quantum mechanics the potentials themselves may be
significant and can determine the dynamics of a charged particle in a
spacetime region where the fields vanish (see, e.g., \cite{Ahar,Pesh}).
In such a situation it is possible that the potential and the field
strength of one eletromagnetic field might couple to the field and the
potential of the other, so that the coupling would give rise to some kind of
observable effects on the particle.
Although dynamical effects on the charged particle by the opposing
electromagnetic fields are cancelled, the presence of potentials may
produce significant results, since one could consider products like
$\phi{\bf E}$ as physical observables. If such terms produce observable
effects on a charged particle, then in the above example, both particles
$A$ and $B$ would give rise to the same dynamical effects on a charged
particle. In the next section it is shown that we can construct such a
model in terms of differential geometry, and that the resulting
electromagnetic effects can be identified with the gravitational force.

\subsection{Asymmetric connection of the form $\Gamma^\sigma_{\mu\nu} =
\Lambda^\sigma_\mu\Phi_\nu$}

We now consider an asymmetric connection of the form
$\Gamma^\sigma_{\mu\nu}=\Lambda^\sigma_\mu\Phi_\nu$.
The motivation for this form of asymmetric
connection is that it allows a construction of an affine connection in
terms of two electromagnetic fields. The quantity $\Phi_\mu$ will then be
identified with the four-vector potential of one electromagnetic field and
the quantity $\Lambda^\mu_\nu$ with the field strength of the second
opposing field. Since the affine connection does not
behave like a tensor under general coordinate transformations, neither
does $\Phi_\mu$ nor $\Lambda^\mu_\nu$. However, they do form tensors under
the group of linear transformations, as should be the case in
electrodynamics. Therefore we must address the problem of
how to couple two electromagnetic fields to produce a spacetime structure
in such a way that the gravitational field can be identified as a
manifestation of that geometry. When the affine connection is formed
from two electromagnetic fields, its possible effects on the motion
of a particle are genuine physical effects if the electromagnetic field
is viewed as a physical field determined from experiment rather than
a geometrical structure, as discussed above. However, due to the
asymmetry of the connection, these effects cannot
be identified with gravity, since the theory of general relativity
requires a symmetric connection. To meet this requirement, it is necessary
to reduce the physical Ricci tensor, which is formed by two electromagnetic
fields, to a symmetric form.

The affine connection of the particular form
$\Gamma_{\mu\nu}^\sigma=\Lambda^\sigma_\mu\Phi_\nu$ reduces the curvature
tensor $R_{\beta\mu\nu}^\alpha=\partial_\mu\Gamma_{\beta\nu}^\alpha -
\partial_\nu\Gamma_{\beta\mu}^\alpha+\Gamma_{\lambda\mu}^\alpha\Gamma_{\beta
\nu}^\lambda-\Gamma_{\lambda\nu}^\alpha \Gamma_{\beta\mu}^\lambda$ to the
simpler form
\begin{equation}
R_{\beta\mu\nu}^\alpha = \frac{\partial {\left ( \Lambda_\beta^\alpha
\Phi_\nu \right )}}{\partial x^\mu} - \frac{\partial {\left (
\Lambda_\beta^\alpha \Phi_\mu \right )}}{\partial x^\nu}.
\end{equation}
The Ricci tensor becomes
\begin{eqnarray}
R_{\mu\nu}&=&\frac{\partial{\left(\Lambda_\mu^\sigma \Phi_\nu\right)}}
{\partial x^\sigma}-\frac{\partial {\left ( \Lambda_\mu^\sigma
\Phi_\sigma \right)}} {\partial x^\nu}\nonumber\\
&=& \left(\frac{\partial \Phi_\nu}{\partial x^\sigma} - \frac{\partial
\Phi_\sigma}{\partial x^\nu} \right ) \Lambda_\mu^\sigma + \Phi_\nu
\frac{\partial \Lambda_\mu^\sigma}{\partial x^\sigma} -  \Phi_\sigma
\frac{\partial \Lambda_\mu^\sigma}{\partial x^\nu}.
\end{eqnarray}
The Ricci tensor in this form can be reduced to a symmetric form if the
quantities $\Lambda_\mu^\nu$ satisfy, for example, the relation
\begin{equation}
\frac{\partial \Lambda_\mu^\sigma}{\partial x^\nu}=\eta_{\mu\nu\sigma}
\Phi_\nu\Lambda_\mu^\sigma,
\end{equation}
where $\eta_{\mu\nu\sigma}$ are arbitrary functions of the coordinate
variables. The indices $\mu,\nu$ and $\sigma$ indicate that the functions
$\eta_{\mu\nu\sigma}$ are specified independently for each term of the
quantity $\Lambda_\mu^\sigma$. For the trivial case of a constant field,
i.e., $\Lambda^\sigma_\mu=constant$, we have $\eta_{\mu\nu\sigma}\equiv 0$.
A more detailed discussion will be given via another example shortly.
The Ricci tensor then becomes
\begin{equation}
R_{\mu\nu}= \Lambda_\mu^\sigma F_{\sigma\nu},
\end{equation}
where we have defined $F_{\mu\nu}=\partial_\mu\Phi_\nu-\partial_\nu
\Phi_\mu$. It is seen that the reduced form (79) of the Ricci tensor is
symmetric if the quantity $\Lambda_\mu^\nu$ is assumed to be the transpose
of the quantity $F_{\mu\nu}$.

In a new coordinate system, ${x'}^\mu={x'}^\mu(x^\nu)$, the transformed
affine connection is still defined as a product $\Gamma'^\sigma_{\mu\nu} =
\Lambda'^\sigma_\mu\Phi'_\nu$. Since the relation (78) retains the same
form in the new coordinate system, due to the arbitrariness of the function
$\eta$, it is apparent that the relation (79) is form-invariant under
general
coordinate transformations; although it behaves like a tensor only under
the linear group of coordinate transformations. Because the affine
connection $\Gamma_{\mu\nu}^\sigma=\Lambda^\sigma_\mu\Phi_\nu$ behaves like
a tensor only under the linear group, the quantities $\Lambda^\sigma_\mu$
and $\Phi_\nu$ behave like a tensor and a vector, respectively, only
under linear coordinate transformations. However, since the tensor
properties of the quantities $\Lambda^\sigma_\mu$ and $\Phi_\nu$ are not
essential in the following discussion, it is sufficient to assume that
their combination transforms like an affine connection under general
coordinate transformations, without further postulating any particular
properties of transformations for each of them separately.

\subsection{Gravity as a coupling of two electromagnetic fields}

The reduced form of the Ricci tensor suggests that in order to incorporate
it into electromagnetism, the quantity $\Phi_\mu$ should be identified
with the four-vector potential and the quantity $F_{\mu\nu}$ with the field
strength of an electromagnetic field; the quantity $F_{\mu\nu}$ has
been defined in terms of the quantity $\Phi_\mu$ by the familiar relation
in electrodynamics, namely, $F_{\mu\nu}=\partial_\mu\Phi_\nu-\partial_\nu
\Phi_\mu$. In this case, if the reduced form of the Ricci tensor is required
to be symmetric, the quantity $\Lambda^\nu_\mu$ can be identified as the
field opposite to the field $F_{\mu\nu}$. The reduced form of the Ricci
tensor therefore may be used as a counterpart of the energy-momentum tensor
to form field equations for gravitation. It should be emphasised again that
the Ricci tensor in this case is a physical quantity which is formed by two
electromagnetic fields of equal magnitude but opposite direction. Its
effect on the motion of a particle in the fields are physical effects which
require physical laws to describe them. These physical laws will be
identified with the field equations of general relativity, because
according to the classical Lorentz force law two
opposing electromagnetic fields of equal magnitude are considered to have
no classical electromagnetic effects on a charged particle moving in the
coupled fields. In terms of the field strengths, the reduced form (2.46) of
the Ricci tensor takes the explicit form
\begin{equation}
R_{\mu\nu} = \left( \begin{array}{cccc}
E_1^2+E_2^2+E_3^2 & E_3B_2-E_2B_3 & E_1B_3-E_3B_1 & E_2B_1-E_1B_2\\
E_3B_2-E_2B_3 & E_1^2+B_2^2+B_3^2 & E_1E_2-B_1B_2 & E_1E_3-B_1B_3\\
E_1B_3-E_3B_1 & E_1E_2-B_1B_2 & E_2^2+B_1^2+B_3^2 & E_2E_3-B_2B_3\\
E_2B_1-E_1B_2 & E_1E_3-B_1B_3 & E_2E_3-B_2B_3 & E_3^2+B_1^2+B_2^2
\end{array}\right).
\end{equation}
The quantity $R_{\mu\nu}$ is a physical quantity which is covariant only
with respect to the group of linear transformations, and this reflects the
covariant properties of the electromagnetic field under the linear group.
Hence, if a symmetric Ricci tensor is required to describe a gravitational
field then only effects caused by the reduced form (80) of the Ricci
tensor are considered, and therefore possible spacetime structures are
restricted to those determined by it. We assume that the dynamics of a
charged particle in a region of spacetime, whose structure is determined
by the connection $\Lambda^\sigma_\mu\Phi_\nu$ formed by two electromagnetic
fields, is governed by the reduced form of the Ricci tensor. This is
reasonable since the dynamics of a charged particle is not effected
classically by the two opposing electromagnetic fields of equal magnitude.
In order to define lengths of the paths of the particle, and hence to
determine dynamical aspects of the particle in the spirit of general
relativity, a new symmetical metric tensor $g_{\mu\nu}$ is introduced
according to the defining relation $ds^2 = g_{\mu\nu}dx^\mu dx^\nu$.
With the introduction of this symmetrical metric tensor into the restricted
spacetime structure, determined by the reduced form of the Ricci tensor,
it is now possible to construct field equations which have a similar form to
the Einstein field equations of gravitation,
$R_{\mu\nu}+\frac{1}{2}g_{\mu\nu}R=\kappa T_{\mu\nu}$, where $T_{\mu\nu}$
is the energy-momentum tensor. The Ricci tensor in these field equations is
constructed from the new symmetrical metric tensor $g_{\mu\nu}$, and is a
geometrical object which is used to describe geometrically the physical
effect caused by the quantity $R_{\mu\nu}$ in (80). Because the field
equations of general relativity do not require a physical basis for the
affine connection, any physical effect on a particle moving in two coupled
electromagnetic fields wil manifest itself through the form of the
energy-momentum tensor rather than through the geometrical Ricci tensor.
Describing physical effects by a geometrical description is fundamental to
the theory of general relativity,
where the Newtonian force is described by a curved spacetime manifold.

It is interesting to note that if we assume that the effect of the coupling
of two opposing electromagnetic fields can be interpreted as a
gravitational field, then the connection
$\Gamma_{\mu\nu}^\sigma=\Lambda_\mu^\sigma\Phi_\nu$, gives rise to a
geodesic equation of the form
\begin{equation}
\frac{d^2x^\mu}{ds^2}+\left(\Phi_\sigma\frac{dx^\sigma}{ds}\right)
\Lambda_\nu^\mu\frac{dx^\nu}{ds}=0.
\end{equation}
It is known that this equation admits a linear first integral of the form
\begin{equation}
\Phi_\mu\frac{dx^\mu}{ds}=-\frac{q}{m},
\end{equation}
provided the quantities $\Phi_\mu$ satisfy the condition \cite{Eise}
\begin{equation}
\nabla_\mu\Phi_\nu+\nabla_\nu\Phi_\mu=0.
\end{equation}
Here we have set the constant in the first integral equal to $-q/m$ for
convenience. This condition identifies $\Phi_\mu$ as a Killing vector
field, which defines a direction of symmetry along which the motion leaves
the spacetime geometry unchanged. The geodesic equation
then has the form of the Lorentz force law
\begin{equation}
\frac{d^2x^\mu}{ds^2}=\frac{q}{m}\Lambda_\nu^\mu\frac{dx^\nu}{ds}.
\end{equation}
We can interpret this result in the following way. When one of the
electromagnetic fields drives the charged particle according to the laws of
classical electrodynamics, the opposite field resists such motion of the
particle and the {\it resistance} manifests itself as the
mass of the particle via the linear first integral (82).

As an illustration, let us consider the simple situation of two
opposite electric fields associated with particles $A$ and $B$ (see
Eqs.(74) and (75)). First, we must examine whether this system
satisfies the relation (78), i.e.,
$\partial_\nu\Lambda_\mu^\sigma=\eta_{\mu\nu\sigma}\Phi_\nu\Lambda_\mu^
\sigma$. If the potentials and the field strengths of the particles $A$ and
$B$ are specified only by the relations (74) and (75), then this system
of fields does not satisfy the relation (78). By rewriting $\Phi_\mu$ and
$\Lambda_\mu^\nu$ explicitly in matrix form
\begin{equation}
\Phi_\mu=(\phi,0,0,0) \ \ \ \ \mbox{and} \ \ \ \ \Lambda_\mu^\nu = \left(
\begin{array}{cccc}
0&-E_x&-E_y&-E_z\\E_x&0&0&0\\E_y&0&0&0\\E_z&0&0&0 \end{array}\right),
\end{equation}
it is seen that $\eta_{\mu\nu 0}\Phi_\nu\Lambda_\mu^0\equiv 0$ but
$\partial_\nu\Lambda_\mu^0\neq 0$ for $\nu\neq 0$. This results from the
fact that in classical electrodynamics a
stationary charged particle does not produce a magnetic field; a pure
gauge vector potential is insignificant because it does not have any
effect (in the context of the classical electrodynamics) on the dynamics of
a charged particle described by the Lorentz force law. However, in the
present situation, if we take the pure gauge vector potential into account
then the difficulty can be resolved. Instead of the relations (74)
and (75), we now consider the potentials and the field strengths
\begin{eqnarray}
{\bf E}_A &=&\frac{kq}{r^3}{\bf r}, \ \ \ \ \phi_A=-\frac{kq}{r}, \ \ \ \
{\bf B}_A=0, \ \ \ \ {\bf A}_A=\nabla\chi\\
{\bf E}_B &=&-\frac{kq}{r^3}{\bf r}, \ \ \ \ \phi_B=\frac{kq}{r}, \ \ \ \
{\bf B}_B=0, \ \ \ \ {\bf A}_B=-\nabla\chi,
\end{eqnarray}
where $\chi$ is an arbitrary function of the coordinate variables. In this
case since $\Phi_\mu\neq 0$ for all $\mu$, the relation $\partial_\nu
\Lambda^\sigma_\mu=\eta_{\mu\nu\sigma}\Phi_\nu\Lambda^\sigma_\mu$ can be
satisfied by a suitable specification of the function
$\eta_{\mu\nu\sigma}$. The reduced form (80) of the Ricci tensor is
written explicitly in terms of the field strengths of the system as
\begin{equation}
R_{\mu\nu}=\frac{k^2q^2}{r^6}\left(\begin{array}{cccc}
r^2 & 0 & 0 & 0\\0 & x^2 & xy &
xz\\ 0 & xy & y^2 & yz\\
0 & xz & yz & z^2 \end{array}\right).
\end{equation}
If a charged particle is located in the coupled fields of particles $A$ and
$B$ then we assume that its dynamics is influenced by this reduced form of
the Ricci tensor. In terms of the electromagnetic field strengths of the two
opposing fields, the dynamical effect on the charged particle caused by
the quantity (88) is regarded as a non-electromagnetic effect, since as we
discussed earlier, dynamical effects on a charged particle due to the
opposing electromagnetic
fields are cancelled. The problem now is to look for a dynamical equation
to describe the effect of the quantity (88) on the motion of the charged
particle. Since in classical physics there are no other known forces,
besides the gravitational force, we identify this effect with gravity
and formulate the problem in terms of the field equations of general
relativity. The procedure is to geometrise the physical object (88) and
to specify its physical effect by defining an energy-momentum tensor in
which the concept of the mass of a particle is introduced. This is similar
to the case where the charge of a particle is introduced into physics for
the purpose of formulating the electrodynamics of a charged particle (see,
e.g., \cite{Jackson}). We note from (88) that the effect is dominated by
the term $R_{00}$ for regions far from the origin at which the charged
particle is located. This result is analogous to the weak field
approximation in general relativity in which Newton's law of gravitation
is recovered \cite{Misn}.

Having asserted that gravity can be described geometrically in terms of
two coupled electromagnetic fields, it seems natural to pose the
question as to whether the connection $\Lambda^\sigma_\mu\Phi_\nu$
itself also produces physical effects. It is observed that
the Ricci tensor (77) can be put in the form
\begin{equation}
R_{\mu\nu}  = \frac{\partial J_{\mu\nu}^\sigma}{\partial x^\sigma},
\end{equation}
where the quantities $J_{\mu\nu}^\sigma$ are defined by
\begin{equation}
J_{\mu\nu}^\sigma = \Lambda_\mu^\sigma \Phi_\nu - \delta_\nu^\sigma
\Lambda_\mu^\lambda \Phi_\lambda.
\end{equation}
In this case the equation $R_{\mu\nu}=0$ results in conservation laws.
For example, the energy density component $R_{00}$ leads to an equation of
continuity formed by the potential and the field strength, i.e.,
\begin{equation}
\frac{\partial ({\bf A}.{\bf E})}{\partial t} + \nabla.(\phi{\bf E})=0,
\end{equation}
where $\Phi_\nu =(\phi,-{\bf A})$, and ${\bf E}$ is the corresponding
electromagnetic field strength of the opposing field.

To conclude this section we remark that in the case where two
electromagnetic fields do not cancel out completely, it is possible to
resolve the fields at each point in a region of spacetime into a pair of
two opposing electromagnetic fields of equal magnitude, and a `residual'
field which acts as a single `conventional' electromagnetic field;
whence, we have the situation where a charged particle is considered to
move under the influence of a combined gravitational and electromagnetic
field. If we denote by $\Lambda_\mu^\sigma\Omega_\nu$ the connection formed
by the opposing fields, where $\Omega_\nu$ represents the
four-potential of one of the two opposing fields, and $\Phi_\mu$ is the
four-potential of the single electromagnetic field, then the change
$\delta V^\mu$ in the components of vector $V^\mu$ under an infinitesimal
parallel displacement is given by \cite{Eise}
\begin{eqnarray}
\delta V^\mu&=&-(\Lambda_\alpha^\mu\Omega_\nu V^\alpha + \beta\Phi_\nu
V^\mu)dx^\nu\nonumber\\
&=&-(\Lambda_\alpha^\mu\Omega_\nu+\beta\delta^\mu\alpha\Phi_\nu)
V^\alpha)dx^\nu.
\end{eqnarray}
It is seen that the total connection is identified with the quantities
$\Gamma_{\mu\nu}^\sigma=\Lambda_\mu^\sigma\Omega_\nu+\beta\delta^\sigma_\mu
\Phi_\nu$. The covariant derivative of a tensor now takes the form
\begin{eqnarray}
\nabla_\sigma A^{\mu_1...\mu_m}_{\nu_1...\nu_n} = \frac{\partial
A_{\nu_1...\nu_n}^{\mu_1...\mu_m}}{\partial x^\sigma} +
\Gamma_{\lambda_i\sigma}^{\mu_i}A^{\mu_1...\lambda_i...\mu_m}_{\nu_1
...\nu_n} &-& \Gamma^{\lambda_i}_{\nu_i\sigma}A^{\mu_1...\mu_m}_{\nu_1...
\lambda_i...\nu_n}\nonumber\\
&+&(m-n)\beta\Phi_\sigma A^{\mu_1...\mu_m}_{\nu_1...\nu_n}.
\end{eqnarray}
The curvature tensor can now be written as
\begin{equation}
R^\alpha_{\beta\mu\nu}=\frac{\partial (\Lambda_\beta^\alpha\Omega_\nu)}
{\partial x^\mu} - \frac{\partial (\Lambda^\alpha_\beta\Omega_\mu)}
{\partial x^\nu} + \beta\left[\frac{\partial (\delta^\alpha_\beta\Phi_\nu)}
{\partial x^\mu} - \frac{\partial (\delta^\alpha_\beta\Phi_\mu)}{\partial
x^\nu}\right],
\end{equation}
and the Ricci tensor reduces to a simple form
\begin{equation}
R_{\mu\nu}=\frac{\partial (\Lambda_\mu^\sigma\Omega_\nu)}{\partial x^\mu}-
\frac{\partial(\Lambda_\mu^\sigma\Omega_\mu)}{\partial x^\nu} + \beta\left(
\frac{\partial \Phi_\nu}{\partial x^\mu}-\frac{\partial \Phi_\mu}{\partial
x^\nu}\right).
\end{equation}
If the parts formed by the first and second terms on the right of this Ricci
tensor is reduced to a symmetric form, for example, by utilising the
relation $\partial_\nu\Phi^\sigma_\mu =
\eta_{\mu\nu\sigma}\Omega_\nu\Lambda^\sigma_\mu$,
then it can be used to describe both the gravitational and electromagnetic
field, whence gravitation can be incorporated consistently into the
electromagnetic spacetime structure.

\subsection{Strong interaction as a coupling of two strong fields}

At short range we have demonstrated that it is possible to describe the
strong interaction by the field equations of general relativity. We now
look for a mechanism responsible for this interaction. We conjecture that
it is a spacetime structure that can be formulated in terms of differential
geometry. As in the case of gravity, which may be thought of as the coupling
of two electromagnetic fields, we postulate an affine
connection of the form
$\Gamma^\sigma_{\mu\nu}=\Lambda^\sigma_\mu\Phi_\nu$. However, the relation
between the quantities $\Lambda^\sigma_\mu$ and $\Phi_\nu$ is now given by
\begin{equation}
\Phi_\sigma\frac{\partial \Lambda^\sigma_\mu}{\partial x^\nu} =-
C^{\alpha\beta}_{\nu\sigma}\Phi_\alpha\Phi_\beta\Lambda^\sigma_\mu,
\end{equation}
where the quantities $C^{\alpha\beta}_{\nu\sigma}$ are arbitrary, but
are assumed antisymmetric with respect to the subscript indices. The
quantity $\Phi_\mu$ is identified with the potential of the strong
interaction, and the quantity $\Lambda^\nu_\mu$ with the field strength.
We further assume antisymmetry of the quantities
$\partial_\nu\Lambda^\sigma_\mu$, which is obtained by imposing the
condition
$\partial_\nu\Lambda^\sigma_\mu+\partial_\sigma\Lambda^\nu_\mu=0$. With
these assumptions, the Ricci tensor (77) then takes the form
\begin{eqnarray}
R_{\mu\nu}&=&\left(\frac{\partial \Phi_\nu}{\partial x^\sigma}-
\frac{\partial \Phi_\sigma}{\partial x^\nu} + C^{\alpha\beta}_{\nu\sigma}
\Phi_\alpha\Phi_\beta\right)\Lambda^\sigma_\mu\nonumber\\
&=&F_{\sigma\nu}\Lambda^\sigma_\mu,
\end{eqnarray}
where the quantity $F_{\mu\nu}$ is defined by the relation
\begin{equation}
F_{\mu\nu}=\frac{\partial \Phi_\nu}{\partial x^\mu} -\frac{\partial
\Phi_\mu}{\partial x^\nu} + C^{\alpha\beta}_{\mu\nu}\Phi_\alpha\Phi_\beta.
\end{equation}
In this form the quantities $F_{\mu\nu}$ can be regarded as a tensor field
for the strong interaction. As with the gravitational field, all strong
sources produce
attractive strong fields, whence it is possible to identify the quantities
$\Lambda^\sigma_\mu$ with the strong field $F_{\mu\sigma}$. Since the
product of two antisymmetric tensors results in a symmetric tensor, the
Ricci tensor has been reduced to a symmetric form, so that the field
equations of general relativity can be applied. As with the case of the
gravitational field, the reduced form (97) of the Ricci tensor behaves
like a tensor only under the linear group of coordinate transformations.
Similarly, the reduced form (97) of the Ricci tensor (77) must be
geometrised in order to investigate the dynamics of a particle under its
influence. This can be carried out by introducing a metric
onto the spacetime manifold, as in the case of a charged particle under the
influence of two coupled electromagnetic fields. The field equations that
govern the metric tensor are assumed to take the form of the field
equations of general relativity, since it has already been shown that the
field equations of general relativity admit a line element of a Yukawa
potential. However, there is a fundamental difference between the present
situation and that discussed for two opposing electromagnetic
fields. In the present case, since the field $\Lambda_\mu^\nu$ is considered
to be identical to the field $F_{\mu\nu}$, the effect of coupling of the two
fields $\Lambda_\mu^\nu$ and $F_{\mu\nu}$, which gives rise to the strong
interaction is additive rather than subtractive as in the case of two
opposing electromagnetic fields.

\section*{Acknowledgements}
I would like to thank Dr. M J Morgan for constructive comments. I also
acknowledge the financial support of an APA Research Award.

\end{document}